\documentclass[aip,pop,a4paper]{revtex4-1}

\usepackage[utf8]{inputenc}
\usepackage{graphicx}
\usepackage{color}
\usepackage{float}
\usepackage{amsmath}
\usepackage{amsthm}
\usepackage{amsfonts}
\usepackage{amsbsy}
\usepackage{amssymb}
\usepackage{amscd}
\usepackage{bm}
\usepackage{xcolor}
\usepackage{hyperref}
\usepackage{subfig}
\usepackage{verbatim}

\newcommand{\e}{\textrm{e}}
\newcommand{\dint}{\textrm{d}}
\newcommand{\im}{\textrm{i}}

\usepackage[colorinlistoftodos]{todonotes}
\begin{document}

\title{Transition from weak to strong turbulence in magnetized plasmas}

\author{Vasil Bratanov} 
\author{Swadesh Mahajan}
\author{David Hatch}
\affiliation{Institute for Fusion Studies, The University of Texas at Austin, Austin, Texas 78712}

\begin{abstract}
\noindent The scaling of turbulent heat flux with respect to electrostatic potential is examined in the framework of a reduced ($4$D) kinetic system describing electrostatic turbulence in magnetized plasmas excited by the ion temperature gradient instability. Numerical simulations were instigated by, and tested the predictions of generic renormalized turbulence models like the $2$D fluid model for electrostatic turbulence [Y.~Z.~Zhang and S.~M.~Mahajan, Phys.~Fluids B 5 (7), pp.~2000 (1993)]. A fundamental, perhaps, universal  result of this theory-simulation combination is the demonstration that there exist two distinct asymptotic states (that can be classified as Weak turbulence (WT) and Strong turbulence (ST) states) where the turbulent diffusivity $Q$ scales quite differently with the strength of turbulence measured by the electrostatic energy $\|\phi\|^2$. In the case of WT $Q \propto \|\phi\|^2$, while in ST $Q$ has a weaker dependence on the electrostatic energy and scales as $\|\phi\|$.
\end{abstract}

\maketitle

\section{Introduction}\label{intro}
\noindent This paper seeks an answer to a fundamental and generic question on the nature of turbulence - How does ``turbulent diffusion'' (some appropriate measure thereof) - one of the most important characteristics of a turbulent state - scale with turbulent energy? There may not be a unique and universal answer for all varieties of turbulence, but is there some partial universality that pertains for some sufficiently broad class of turbulent phenomena? To begin the quest for an answer, we will study here in some depth a particular but very important system - electrostatic turbulence in a magnetized plasma - often observed, both in the laboratory and in astrophysical settings.\\
We will attempt to find the answer through a combination of analytical reasoning (inspired from a model of renormalized turbulence theory) and numerical simulations. The latter will be based on a reduced gyrokinetic model (concentrating on the ion temperature gradient (ITG) driven instability) that will be described in Section III. But to provide a context for the problem (and clues towards a possible solution), we will begin by discussing, in some detail, the essential content of a specific, though, typical $2$-dimensional fluid model constructed  to investigate electrostatic plasma turbulence \cite{Zhang1}. The relatively simple  model \cite{Zhang1} is based on a continuity equation for the electron density and the quasi-neutrality constraint coupled to an equation for the generalized enstrophy $\Psi = \ln(n) - \Delta_{\perp}\Phi$ where $\Delta_{\perp} = \partial^2/\partial x^2 + \partial^2/\partial y^2$. $\Psi$ is an inviscid constant of motion constructed from the electron number density $n$, and the electrostatic potential $\Phi$.\\
When ion parallel motion is neglected, the turbulent dynamics of this system (immersed in a constant guiding magnetic field of magnitude ${B}$) becomes essentially two-dimensional and is modelled by a fluid-like evolution equation for $\Psi$ ($c$ is the speed of light)
\begin{equation}\label{eq_gen_enstrophy} 
\left( \frac{\partial}{\partial t} + \frac{c}{B}(\mathbf{b}\times\nabla\Phi)\cdot\nabla - \mu\Delta \right)\Psi = 0 \mbox{.}
\end{equation}
Note that in this model, the electrostatic potential $\Phi$ consists of an equilibrium part $\phi_0$ and a fluctuating part $\phi$. The unit vector $\mathbf{b} = \mathbf B/B$ is perpendicular to the plane of the enstrophy dynamics described by Eq.~\eqref{eq_gen_enstrophy}. The positive coefficient $\mu$ plays the role of a ``classical" viscosity. One readily sees that in the inviscid case ($\mu$=0), $\Psi$ is constant along the orbit (dictated by the $\mathbf{E}\times\mathbf{B}$ motion). It is worth pointing out that the model  \cite{Zhang1} can be reduced to other well-known $2$D fluid models for plasma turbulence, for instance, the one embodied in the Hasegawa-Mima equation.\cite{Hasegawa1, Hasegawa2} Key quantities of interest here are $N$-body correlation functions of the form $\langle\Psi(\mathbf{r}_1,t)\Psi(\mathbf{r}_2,t)\cdots\Psi(\mathbf{r}_N,t)\rangle$.\\
Although we will be eventually dealing with systems that reach a turbulent state by means of some internally-driven instabilities 
(drift class of instabilities in confined plasmas), this idealized model has neither an internal instability nor an explicit forcing term. Instead, the  drive is simulated by considering $\phi$ (the fluctuating part of the electrostatic potential) as an external field that is stochastic with a Gaussian distribution in time. With the aid of the Gaussian assumption and the use of the Novikov theorem one can derive an evolution equation for the ensemble average of the generalized enstrophy, i.e.,
\begin{equation}\label{eq_1psi_av}
\left( \frac{\partial}{\partial t} + (\mathbf{b}\times\nabla_i\phi_0(\mathbf{r}_i,t))\cdot\nabla_i - \mu\Delta_i - \nabla_i\cdot\mathbf{D}_{ii}\cdot\nabla_i \right)\langle \Psi(\mathbf{r}_i,t)\rangle = 0 \mbox{,}
\end{equation}
where the tensor $\mathbf{D}_{ii}$ can be interpreted as representing a generalized turbulent ``diffusion" (in analogy with the viscous diffusion proportional to the scalar coefficient $\mu$); its detailed expression is 
\begin{equation}\label{eq_Dij_real}
\mathbf{D}_{ij} := \intop_{t_0}^{t}\langle (\mathbf{b}\times\nabla_i\phi(\mathbf{r}_i(t),t))\otimes
(\mathbf{b}\times\nabla_j\phi(\mathbf{r}_j(t'),t')) \rangle \dint t' \mbox{,}
\end{equation}
where $\otimes$ denotes the tensor product of two vectors. Analogous equations can be derived also for the higher-order correlation functions and their exact form is given in Ref.~\cite{Zhang1}. For mathematical convenience, one can express the diffusion tensor via auxiliary quantities in Fourier space as
\begin{equation}\label{eq_Dij_F}
\mathbf{D}_{ij} = \intop (\mathbf{b}\times\mathbf{k})\otimes(\mathbf{b}\times\mathbf{k})\Pi(\mathbf{k})
\exp(\im\mathbf{k}\cdot(\mathbf{r}_i - \mathbf{r}_j))\dint^2\mathbf{k} \mbox{,}
\end{equation}
where $\Pi(\mathbf{k})$ depends on the averaged one-particle Green's function and the fluctuation amplitude $|\phi|^2$ as
\begin{equation}\label{eq_cap_Pi}
\Pi(\mathbf{k}) = \Re\left( \intop\frac{\im\langle|\phi(\omega,\mathbf{k})|^2\rangle}{\omega - \mathbf{k}\cdot(\mathbf{b}\times\nabla\phi_0) - \im\mathbf{k}\cdot\mathbf{D}_{jj} \cdot \mathbf{k}}\dint\omega \right) \mbox{.}
\end{equation}
Notice that the effects of turbulent fields on the dynamics appear only through $\Pi(\mathbf{k})$ - naturally it is the quantity that must be examined to extract information about the turbulent state. We note:

1) Since both the right and the left side of Eq.~(\ref{eq_Dij_F}) depend on the turbulence diffusion tensor, it constitutes an implicit equation for determining $ \mathbf{D}_{ij}$. 

2)The turbulent diffusion tensor ($ \mathbf{D}_{jj}$) modifies the propagator in Eq.~(\ref{eq_cap_Pi}) from its linear form $\omega - \mathbf{k}\cdot(\mathbf{b}\times\nabla\phi_0)$. This is a general qualitative feature of most renormalized turbulence theories
 
3) In addition, the integrand in $\Pi(\mathbf{k})$ is directly proportional to $|\phi|^2$ (that can be viewed as a measure of the total electrostatic energy) implying that  $\mathbf{D}_{ij}$ will necessarily depend on $|\phi|^2$. 

4) The form of the resonant denominator in Eq.~(\ref{eq_cap_Pi}) suggests two asymptotic regimes. Since the goal of our theoretical analysis is to advance qualitative understanding (so that we could interpret the results of simulations with greater confidence), we shall ignore the Doppler correction $\mathbf{k}\cdot(\mathbf{b}\times\nabla\phi_0)$ to the real frequency $\omega$ . We must inform the reader that this very term, originating in differential plasma rotation, is a major mechanism for turbulence suppression in tokamkas\cite{Terry, Hatch2, Boedo, Schaffner}.  Once the Doppler shift is dropped, the ratio between $\omega$ and $|\mathbf{k}\cdot\mathbf{D}_{jj}\cdot\mathbf{k}|$ will determine the relative ``strength" of the linear and the turbulent contributions. Notice that neglecting the Doppler shift is for simplicity alone; one could readily deal with it if the equilibrium flow shear is significant.\\
The resonant propagator has two obvious asymptotic limits:

a) When $\omega \gg |\mathbf{k}\cdot\mathbf{D}_{jj}\cdot\mathbf{k}|$ ( this may, indeed, be taken as the definition of weak turbulence), the contribution of turbulent diffusion to Eq.~(\ref{eq_cap_Pi}) may be neglected. Under those circumstances, $\Pi(\mathbf{k})$, and therefore, the measure of turbulent diffusion $D$ will scale, schematically, as $\sim |\phi|^2$. It is worth remarking that for plasma turbulence (built around fluctuations that have a real frequency), a weak turbulence state is totally legitimate. In this regard plasma turbulence does differ, qualitatively, from the conventional Navier Stokes turbulence. 

b) In the opposite  extreme limit when $\omega \ll |\mathbf{k}\cdot\mathbf{D}_{jj}\cdot\mathbf{k}|$, the real frequency can be neglected, and the propagator is dominated by turbulent diffusion.  Equation~(\ref{eq_Dij_F}), then, is schematically equivalent to $D \sim |\phi|^2/D$ yielding the scaling $D \sim |\phi|$. Naturally this is the regime of super-strong turbulence; the system is left with little memory of the linear regime, the Green's function is set, primarily, by turbulent fluctuations. For the rest of the paper these two asymptotic states will be referred to as WT and ST, respectively.

5) It is, of course, possible that turbulence in real physical systems may not ever approach the ST state.\\
At this point a clarification regarding the terminology (used in this paper) is required in order to avoid possible confusion. In the MHD literature, the terms `weak' and `strong' turbulence  were introduced in a couple of seminal papers by Goldreich and Sridhar.\cite{Sridhar, Goldreich}. The weak turbulence is defined as a state with small nonlinear interaction between well-defined linear waves. This is closely related to the parallel streaming time being much smaller than the nonlinear time with the latter being defined as the time that it takes for nonlinear processes to transfer a significant amount of energy between different modes. Strong turbulence in Refs.~\cite{Sridhar, Goldreich}, on the other hand, is a state in which those two times are comparable. Such an equality is referred to as `critical balance'. A similar analysis has been done also for gyrokinetic systems in toroidal geometry\cite{Barnes}; the critical balance, then, manifests as a dynamic adjustment of the nonlinear spectrum such that the characteristic parallel wavenumber, computed as an average of all parallel wavenumbers weighted by the nonlinear energy, matches the characteristic perpendicular wavenumber. Critical balance has already been investigated and confirmed for the model that we consider in this paper.\cite{Hatch1} However, the notion of critical balance is different from what we consider in this paper; here we compare a purely linear frequency $\omega$ to a nonlinear one that arises from the turbulent diffusion. We shall use for the rest of this paper the terms `weak' and `strong' turbulence because they arise naturally based on the definition we give and the limits considered in a) and b), and the reader should bear in mind that the meaning behind them is different from that in Refs.~\cite{Sridhar, Goldreich}. It should be emphasized that there could potentially be some relation or even equivalency in some way between our definitions of WT and ST, and those that prevail in MHD turbulence. The arguments (as well as the salient results) presented in this paper, however, do not depend on the existence of such a relation and an investigation of that issue will be the object of future work.\\
The most important prediction of the preceding analysis is that the amplitude dependence of the turbulent diffusion (an appropriate scalar measure  of the diffusion tensor) changes from $D\sim |\phi|^2$ in WT to $D\sim |\phi|$ in ST. We should, however, remember that in our model, the turbulent field was externally specified while in a typical drift-wave system (the main object of the current investigation through numerical simulations), the turbulence is generated by an internally driven instability. Therefore, we have to construct an appropriate translation methodology to compare simulations with qualitative analytic predictions.
\section{From analytic theory to numerical simulations}\label{2}
\noindent As mentioned earlier, our simulations will be based on a reduced gyrokinetic (RGK) model considerably more encompassing than the analytical model of Ref.~\cite{Zhang1}. The exact mathematical formulas in Ref.~\cite{Zhang1} cannot be directly adopted when analyzing simulation results. One  hopes that the conceptual framework (summarized above) will guide our understanding of simulation results presented in Sections \ref{or_dna}, \ref{coll_scan} and \ref{mod_dna}. In fact we hope to test/verify the analytical predictions for the  $|\phi|$ scaling of $D$.\\
The reduced gyrokinetic model (in a slab geometry) will be quantitatively outlined in Section \ref{dna}. {\it We will concentrate on a 
temperature-gradient-driven instability whose principal nonlinear manifestation will be the enhanced thermal diffusion; the heat diffusivity $Q$ will be taken as a proxy and a measure for the turbulent diffusion tensor of the model theory.}\\
The transition from weak to strong turbulence will be affected by increasing the normalized temperature gradient $\omega_T$ which in turn boosts the growth rate of the acting instability. All other parameters are kept constant. The stronger gradients then result in higher saturation amplitudes for both electrostatic potential $|\phi|$ and heat flux $Q$.\\
However, this procedure of driving the system toward an ST state, turns out to be problematic but very interesting and revealing. Analysis of the corresponding numerical simulations failed to show any transition for the exponent in the relation between $Q$ and $|\phi|$ as the system was driven harder and harder (by increasing $\omega_T$) boosting the electrostatic potential by several orders of magnitude. A single slope, characterized by an exponent  around $1.71$, was observed (these simulations will be described in detail in Sec.~\ref{or_dna}). The exponent  lies in between the asymptotic values of $1$ and $2$; closer to  $2$, the regime of WT.\\
At first sight, these simulations seem to spell disaster for the analytical model in Ref.~\cite{Zhang1}. Instead, the simulations  actually fortified the model by forcing a more profound, though obvious, reexamination of the resonance denominator. Let us remind ourselves that the regimes of weak and strong turbulence are distinguished by the ratio $\omega/ |\mathbf{k}\cdot\mathbf{D}_{jj}\cdot\mathbf{k}|$ and not by the individual magnitude of either term. It just so happens that for $\omega_T$-driven turbulence, an increase in $\omega_T$ not only pushes $|\phi|$ (and the turbulent diffusion) up, it also increases $\omega$, the real frequency of the dominant mode - the ITG instability, proportionately to $\omega_T$. Thus, as $\omega_T$ is raised, the system is stuck at the same ratio $\omega/ |\mathbf{k}\cdot\mathbf{D}_{jj}\cdot\mathbf{k}|$, and no transition should be expected even when the turbulence amplitudes are increased by orders of magnitude.\\
It becomes equally evident that in order to observe the (WT-ST) transition as some parameter (like $\omega_T$) is stepped up to amplify the electrostatic potential (by inducing a larger growth rate $\gamma$), the real frequency $\omega$ must not be affected much, i.e, $\gamma$ and $\omega$ must be disconnected if the nature of turbulence is to change with levels of turbulent fields. In fact, when such a decoupling is created in simulations, we do see the transition from WT ($Q \sim |\phi|^2$) to ST ($Q \sim |\phi|$) as the turbulence builds up.
\section{Reduced gyrokinetic model}\label{dna}
\noindent The numerical results described in this work are obtained with the aid of a simplified kinetic model (described in detail in  Refs.~\cite{Hatch0, Hatch1}) for the one-particle distribution function from which higher-order moments can then be constructed. The physical setup is that of a fully-ionized plasma (singly-charged ions) in a strong, constant magnetic field $\mathbf{B} = B\mathbf{e}_z$, i.e., there is no magnetic shear. Due to the specifics of the problem the computational domain is square in the $x$- and $y$-direction and elongated along the $z$-axis with periodic boundary conditions in all directions for numerical convenience. In addition, the density and temperature gradients are present only along the $x$-direction and are treated in the so-called local-gradient approximation (to be quantitatively defined later). The temperature gradient, in particular, is the one that has the potential to give rise to linear instabilities that drive the turbulence. A fully kinetic description of the plasma dynamics must involve the evolution of the distribution function in a $6$-dimensional phase space spanned by the three spatial and three velocity co-ordinates. However, in strongly magnetized plasmas, the gyro-motion is usually much faster than the turbulent dynamics. An average over the fast scale leaves us with the so-called gyrokinetic model (for a modern introduction into gyrokinetics see Ref.~\cite{Brizard}) where there are only two velocity co-ordinates: parallel and perpendicular to the guiding magnetic field, i.e., $v_{||}$ and $v_{\perp}$. The latter is closely related to the magnetic moment $\mu = m_i v_{\perp}^2/(2B)$ of the gyrating particle. The model utilized in this paper assumes an adiabatic electron response and follows the so-called `$\delta f$-approach' where the distribution function is decomposed into a Maxwellian $F_{0}(v_{\parallel},\mu)$ and a perturbation $f$ the magnitude of which is much smaller than $F_{0}$. The equation of motion for the perturbation $f(\mathbf{r},v_{\parallel},v_{\perp},t)$ in gyro-center co-ordinates is then
\begin{multline}\label{eq_dna_real}
\qquad\frac{\partial f}{\partial t} = -\left( \omega_n + \omega_T\left( v_{\parallel}^2 + \mu - \frac{3}{2} \right) \right)F_{0}(v_{\parallel},\mu)
\frac{\partial}{\partial y}\overline{\phi} - \sqrt{2}v_{\parallel}\frac{\partial}{\partial z}\left( \overline{f} - 
F_{0}\overline{\phi} \right) +\\
+ \nu C\left(f\right) - \left(\frac{\partial\overline{\phi}}{\partial y}\right)\frac{\partial f}{\partial x} + 
\left(\frac{\partial\overline{\phi}}{\partial x}\right)\frac{\partial f}{\partial y}\mbox{,}\qquad\qquad\qquad\qquad\qquad\qquad
\end{multline}
where the overbar designates gyro-averaging, i.e., integrating the corresponding variable over the gyro-angle $\theta$: $1/\pi\intop_{0}^{2\pi}\dint\theta$. In the electrostatic approximation considered here, Eq.~(\ref{eq_dna_real}) has to be supplemented with the Poisson equation that provides another relation between $f$ and $\phi$. If the Debye length is neglected it reads
\begin{equation}\label{eq_phi_real}
(1 + \tau)\phi - \overline{\overline{\phi}} = \intop_{-\infty}^{+\infty}\intop_{0}^{+\infty}\overline{f}\dint v_{\parallel}\dint\mu
\mbox{,}
\end{equation}
where $\tau$ is the ratio between the ion and electron background temperatures, i.e., $\tau = T_{i0}/T_{e0}$ (set here to unity), while $\Gamma_0(k_{\perp}^2) = \e^{-k_{\perp}^2}I_0(k_{\perp}^2)$ and $I_0$ stands for the modified Bessel function of order $0$. Since we use normal Cartesian co-ordinates the perpendicular wavenumber is simply $k_{\perp}^2 = k_x^2 + k_y^2$. The quantities in the above equations are normalized based on the background ion density $n_{i0} := n_i(x_0)$ taken at a fixed reference point $x_0$, the density gradient scale length $L_d$, the thermal ion velocity $v_{th,i}$, the elementary electric charge $e$, the background electron temperature $T_{e0} = T_{e}(x_0)$ and the thermal ion Larmor radius defined as $\rho_i = m_i v_{th,i}/(eB)$. The physical, i.e., dimensional, quantities relate to the ones in Eq.~(\ref{eq_dna}) in a straightforward way: $f\rho_i n_{i0}/(L_d v_{th,i})$ stands for the one-particle ion distribution function, $\overline{\hat{\phi}}\rho_i T_{e0}/(L_d e)$ is the gyro-averaged electrostatic potential while $\nu v_{th,i}/L_d$ equals the collision frequency. With that normalization the background distribution is given by $F_0(v_{\parallel},\mu) = \pi^{-3/2}\exp(-v_{\parallel}^2 - \mu)$. Temperature and density gradients are defined, respectively, as $\omega_T = L_d(\dint T_i/\dint x)/T_i |_{x = x_0}$ and $\omega_d = L_d(\dint n_i/\dint x)/n_i |_{x = x_0}$. In the local-gradient approximation those are assumed to be constant. Over the thin, flux-tube-like simulation domain the variation of temperature and density is assumed to be small enough as to be neglected, i.e.,
\begin{gather}
n_i(x) \approx n_i(x_0)(1 + \omega_d (x - x_0)/L_d) \approx n_i(x_0) = n_{i0} = \textrm{const} \\
T_i(x) \approx T_i(x_0)(1 + \omega_T (x - x_0)/L_d) \approx T_i(x_0) = T_{i0} = \textrm{const} \mbox{.}
\end{gather}
This is justifiable when the length-scale $L_d$ over which the temperature and density vary is much larger than the extent of the plasma in the $x$-direction, i.e., $\max(|x-x_0|) \ll L_d$. However, the derivatives of density and temperature shall still be taken into account (via assuming constant $\omega_d$ and $\omega_T$) when they appear separately in different terms.\\
The system can be simplified even further if one assumes that the distribution is a Maxwelian with respect to $v_{\perp}$, i.e., $f \propto \e^{-\mu}/\pi$. Then a $\mu$ integration leaves us with a $4$-dimensional reduced gyrokinetic model (RGK). Rudimentary finite Larmour radius effects are preserved so that the approximation is well-justified for perpendicular spacial scales of $k_{\perp}\rho_i \lesssim 1$\cite{Dorland1} and retains important kinetic effects, e.g., phase mixing and Landau damping. The only remaining (parallel) velocity dependence is handled by expanding the Fourier-transformed distribution function in terms of the orthonormal set of Weber-Hermite polynomials,
\begin{equation}\label{expansion}
f(\mathbf{k},v_{||},t) = \sum_{n=0}^{+\infty}\hat{f}_n(\mathbf{k},t) H_n(v_{||})\e^{-v_{||}^2} \mbox{,}
\end{equation}
where $H_n(x) = (n!2^n\sqrt{\pi})^{-1/2}\e^{x^2}(-\dint/\dint x)^n\e^{-x^2}$ is the Hermite polynomial of order $n$. Such an approach has proven to be rather useful in kinetic plasma systems, primarily, because all terms in the expansion decrease exponentially (like the Maxwellian distribution) at large parallel velocities. Since this is the expected large-$v_{||}$ behavior of the exact solution, the Weber-Hermite expansion may constitute an optimal decomposition.\\
The coefficients $\hat{f}_n(\mathbf{k},t)$ in Eq.~(\ref{expansion}) are shown to obey the reduced (normalized) gyrokinetic equation, 
\begin{multline}\label{eq_dna}
\frac{\partial\hat{f}_n(\mathbf{k},t)}{\partial t} = - \nu n\hat{f}_n(\mathbf{k},t) + \frac{\im}{\pi^{1/4}}\left( k_y\left(\omega_T\frac{k_{\perp}}{2} - \omega_d\right)\delta_{n0} - \frac{k_y\omega_T}{\sqrt{2}}\delta_{n2} - k_z\delta_{n1} \right)\overline{\hat{\phi}(\mathbf{k},t)} \quad - \\
- \im k_z\left( \sqrt{n}\hat{f}_{n-1}(\mathbf{k},t) + \sqrt{n+1}\hat{f}_{n+1}(\mathbf{k},t) \right) + 
\sum_{\mathbf{k}'}(k_x' k_y - k_x k_y')\overline{\hat{\phi}(\mathbf{k}',t)}\hat{f}_n(\mathbf{k}-\mathbf{k}',t) \mbox{.}
\end{multline}
The electrostatic potential is directly related to the $0$th Hermite coefficient of the distribution function as
\begin{equation}\label{eq_potential}
\overline{\hat{\phi}(\mathbf{k},t)} = \frac{\pi^{1/4}\e^{-k_{\perp}^2/2}\hat{f}_0(\mathbf{k},t)}{1 + \tau - \Gamma_0(k_{\perp}^2)} \mbox{.}
\end{equation}
It is to be noted that the $z$ co-ordinate, which is aligned with the magnetic field, is normalised over $L_d$ while the normalization factor in the other two perpendicular directions is $\rho_i$. Same applies, of course, to the corresponding wavenumbers. All simulations presented in this work use $48$ Hermite moments and $64$ Fourier modes in all spatial dimensions, resolving from $k_{x,y} = 0.05 - 1.55$ in the perpendicular directions and $k_z = 0.1 - 3.1$ in the parallel direction. Hyperdiffusion terms of order $8$ are used in the perpendicular directions to suppress fluctuations at $k_\perp \rho_i > 1$ beyond which our treatment of perpendicular velocity space (i.e., integrating it out) becomes invalid. Turbulence in slab ITG systems tends to be suppressed by zonal flows \cite{Watanabe, Hatch0}. Hence, we have opted for an ETG-like adiabatic response as well as an additional hyper-viscosity terms acting on the zonal flows alone that reduce their strength to very low levels and allow for the turbulence to develop. Further numerical details can be found in Ref.~\cite{Hatch1}.\\
At this point some comments regarding the physical model and the way in which the system reaches saturation are due. In the local-gradient approximation $\omega_T$ and $\omega_d$ are constant and held fixed, i.e., do not evolve due to turbulence diffusion. Hence, the diffusion increases until it adjusts to the given gradients. The free energy in the temperature gradient, when above a certain threshold, drives a linear instability. Initially the amplitude of the velocity moments grows only in the unstable wavenumber regime. Eventually, the nonlinear terms, which are quadratic with respect to the velocity moments, become comparable to the linear part and cancel the effect of the instability. The constant gradients always provide a source of free energy for the system that is eventually dissipated via collisional processes and turned into heat. The contribution of the nonlinearity to the saturation is indirect by coupling different wavenumbers and transferring free energy to different (linearly stable) modes. A realistic collisional operator has diffusive terms both in $v_{\parallel}$ and $v_{\perp}$. In our reduced model there is only $v_{\parallel}$. Hence, our collision operator involves only the Hermite index $n$ (that corresponds to the resolution in $v_{\parallel}$). Once higher $n$s are excited, i.e., smaller structure in $v_{\parallel}$-space develops, the effect of collisions becomes stronger and free energy is dissipated. A collision operator in $v_{\perp}$-space is approximated by a hyperdiffusion term in $k_{\perp}$. This is justified because full, $5$-dimensional gyrokinetic simulations have shown that small structure in perpendicular space (excited by the nonlinearity) leads to the fluctuations developing also small structure in $v_{\perp}$-space.\\
The numerical solutions of Eq.~(\ref{eq_dna}) analysed in this work are obtained with the aid of the DNA code, which has been used for several basic turbulence studies~\cite{Hatch0, Hatch1, Hatch_NJP}. DNA employs pseudo-spectral methods in all three spatial dimensions. Time advancement is obtained via a $4$th-order Runge-Kutta scheme with an adaptive time step that satisfies the Courant-Friedrichs-Lewy condition. For further technical details of the DNA code on can consult Ref.~\cite{Hatch1}. 
\section{Transition from weak to strong turbulence}\label{or_dna}
\noindent With the use of the Hermite representation, as defined in the previous section, the different coefficients $\hat{f}_n$ in the decomposition of the distribution function are closely related to its moments. Each $\hat{f}_n$ can be written as a linear combination of the first $n$ moments of $\hat{f}$ with respect to $v_{||}$. Hence, Eq.~(\ref{eq_dna}) constitutes a system of infinitely many coupled $1$st-order ordinary differential equations closely related to the infinite hierarchy of moment equations that is well known in kinetic theory. A numerical solution is obtained by truncating this hierarchy at some $n$, i.e., setting $\hat{f}_m \equiv 0$ for all $m > n$. Physical quantities like temperature, pressure or heat flux can then be easily related to the coefficients $\hat{f}_n$. For the heat flux $Q$ the corresponding equation reads
\begin{equation}\label{eq_heat_flux}
Q(t) = -\frac{\pi^{1/4}}{\sqrt{2}}\sum_{\mathbf{k}}\im k_y\e^{-k_{\perp}^2/2}\hat{\phi}(\mathbf{k},t)\hat{f}_2^*(\mathbf{k},t) \mbox{,}
\end{equation}
where the asterisk next to $\hat{f}_n$ denotes complex conjugation. Since turbulence is driven by the temperature gradient, the heat flux is thermodynamically enforced to be positive (i.e. down the gradient) in a statistical sense. The other quantity of interest that we want to relate to the heat flux is the electrostatic potential $\hat{\phi}$.

\begin{figure}[h]
    \centering
    \subfloat[Electrostatic potential]{{\includegraphics[width=8.0cm]{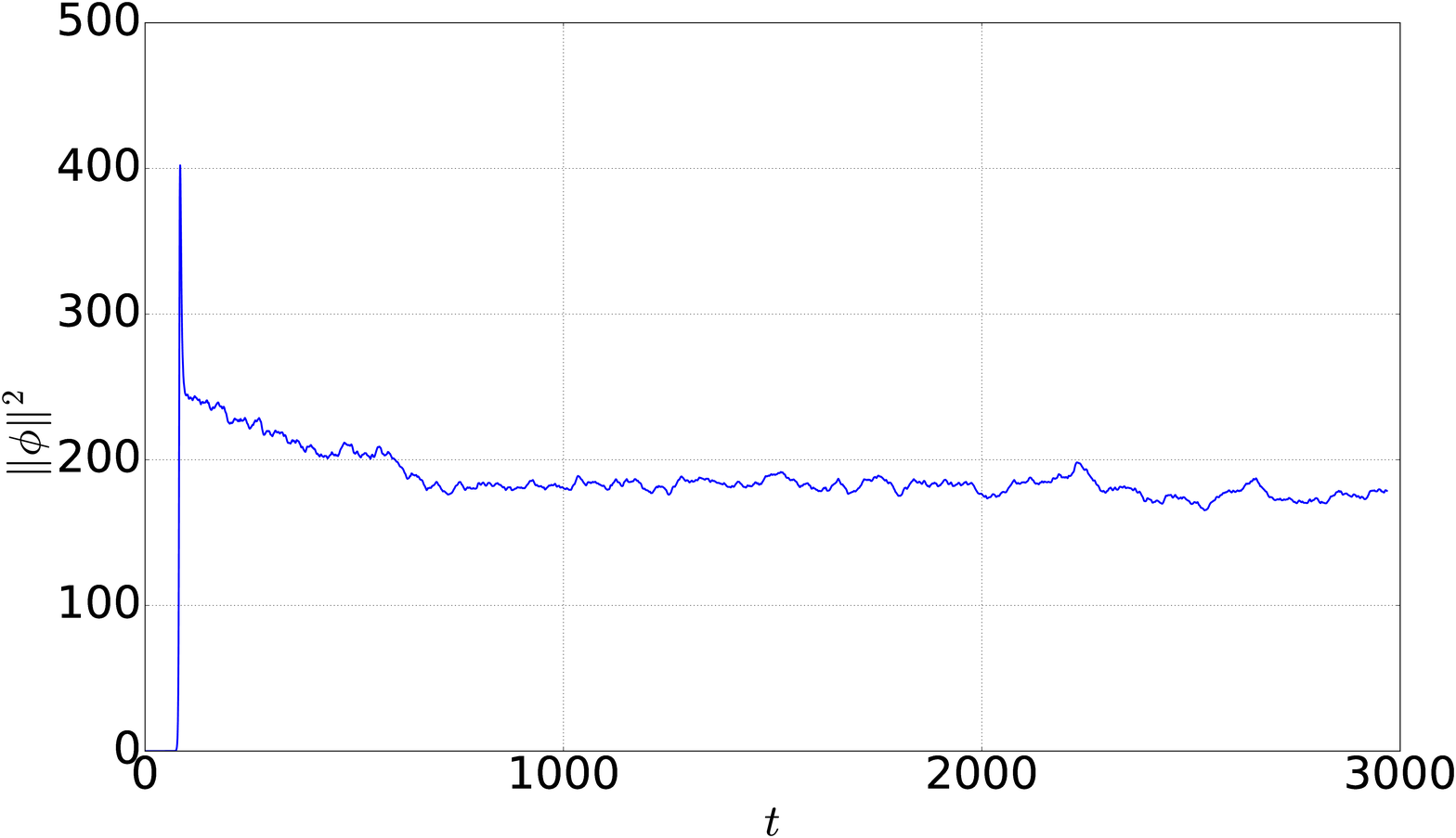} }}
    \subfloat[Heat flux]{{\includegraphics[width=8.0cm]{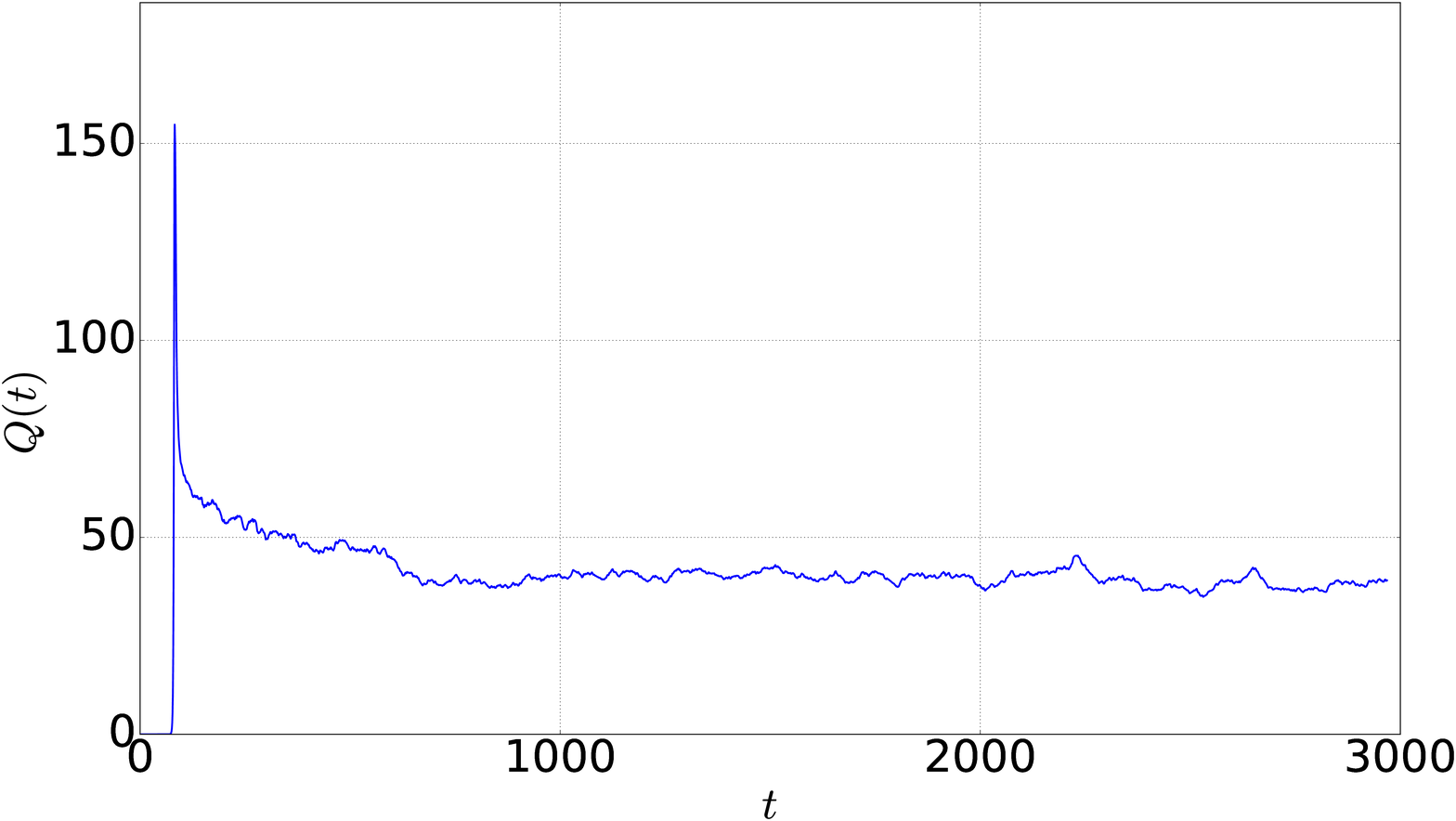} }}
    \caption{Time trace of electrostatic potential (summed over all Fourier modes) and heat flux from a typical numerical simulation with DNA. In normalized units, the inverse temperature and density gradient lengths are set to $\omega_T = 40$ and $\omega_d = 1.0$, respectively, while collision frequency is $\nu = 2.2$.}
    \label{fig_timetrace}%
\end{figure}

\begin{figure}[h]
    \centering
    \includegraphics[width=0.8\textwidth]{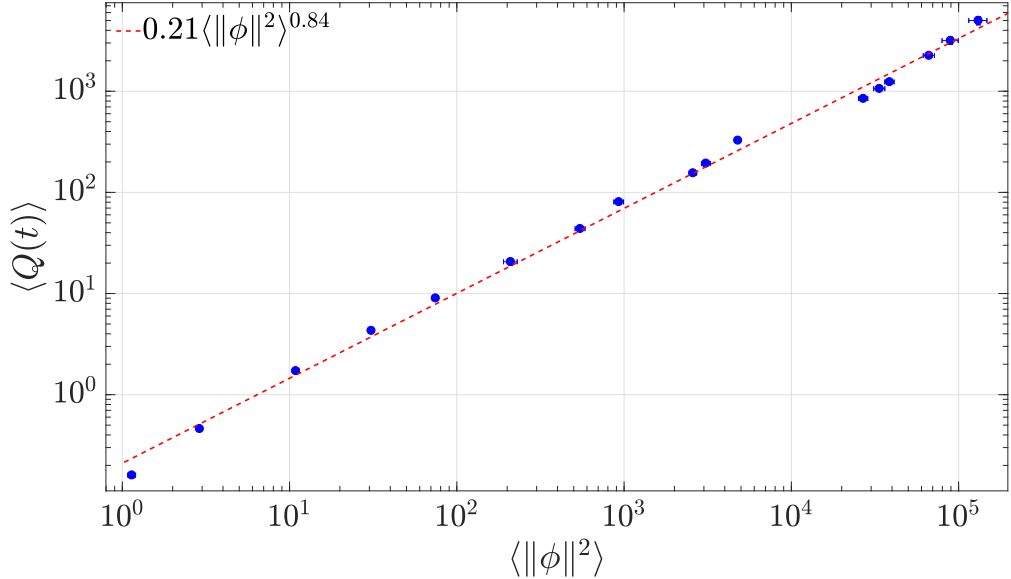}
    \caption{A double logarithmic representation of the dependence of heat flux of the amplitude of the electrostatic potential when the drive, i.e., the temperature gradient, is varied. The different points correspond to different values of $\omega_T$; from left to right $\omega_T = 2.5, 3, 4, 5, 6, 7, 8, 10, 11, 12, 15, 18, 19, 20, 25, 30, 40$.}
    \label{fig_Q_phi_or}
\end{figure}

Since we are interested in the system as a whole, we shall use the $L^2$ norm of $\phi$ as a global measure of the electrostatic potential. Due to the unitarity of the Fourier transform this is equivalent to the sum of the amplitude of $\hat{\phi}$ over all Fourier modes, i.e.,
\begin{equation}
\|\phi\|^2 = \frac{1}{L_x L_y L_z}\intop_{0}^{L_x}\intop_{0}^{L_y}\intop_{0}^{L_z}|\phi(x,y,z,t)|^2\dint x\dint y\dint z = 
\sum_{\mathbf{k}}|\hat{\phi}(\mathbf{k},t)|^2 \mbox{,}
\end{equation}
where $L_x$, $L_y$ and $L_z$ are the lengths of the computational domain in $x$-, $y$- and $z$-direction, respectively. Fig.~\ref{fig_timetrace} shows an example for the time trace 
of electrostatic potential (a) and heat flux (b) from a typical simulation. The displayed case has temperature and density gradients of $w_T = 40$ and $w_d = 1$ while the collision frequency is set to $\nu = 2.2$. This set of parameters leads to linear instabilities (source of the turbulent drive)  in roughly $20$\% of the wave-number spectrum. The initial condition used for this computation had a very small amplitude which resulted in the nearly exponential increase of both $\|\phi\|$ and $Q$ at the beginning of the numerical simulation. After a relatively short transient period the system attains a statistically stationary state. The existence of such a statistically stationary state allows us to effectively substitute the ensemble average used in the analytical calculations in Ref.~\cite{Zhang1} with a time average (designated here also by angular brackets, i.e., $\langle\cdot\rangle$) that is obtained much more easily. Applying that to quantities like the heat flux and the electrostatic potential allows us to calculate a pair of values $(\langle\|\phi\|^2\rangle, \langle Q\rangle)$ for each simulation, i.e., for each parameter set $\omega_T$, $\omega_d$ and $\nu$. The fact that we reach a stationary state only in a statistical sense means that the corresponding quantity still fluctuates around its average. The amount of fluctuation can be considered as a sort of an uncertainty regarding the exact value of the average. It allows us to attach error bars to the data points obtained from nonlinear simulations. More precisely, the error bars in the figures below correspond to one standard deviation.\\
In an attempt to reach the two asymptotic turbulent regimes discussed in Sections \ref{intro} and \ref{2} we vary the temperature gradient. This leads to different saturation levels for the two turbulent quantities of interest and the results from the numerical simulations are summarized in Fig.~\ref{fig_Q_phi_or} where $\langle Q\rangle$ is plotted against $\langle\|\phi\|^2\rangle$ as $\omega_T$ is varied over a considerable range. The temperature gradient steepens from left to right. The blue points denote the simulation results. As expected, both $\langle Q\rangle$ and $\langle\|\phi\|^2\rangle$ increase with $\omega_T$. However, all the points lie on a straight line of definite slope which does not change when going from small to large fluctuation amplitudes.\\
The leftmost data point corresponds to $\omega_T = 2.5$ at which less than $0.2$\% of the Fourier components possess a linear instability. This was the lowest-driven case that we were able to reach numerically such that a turbulent state still develops and the nonlinear interactions are not negligible. On the other hand, the rightmost point was obtained with $\omega_T = 40$ leading to nearly $40$\% of unstable wavenumbers. The double logarithmic representation is chosen in order to highlight a power-law dependence of $\langle Q\rangle$ on $\langle\|\phi\|^2\rangle$ of the type that we conjectured, i.e.,
\begin{equation}\label{eq_Q_form}
\langle Q(t)\rangle = C \left( \sum_{\mathbf{k}}\langle|\hat{\phi}(\mathbf{k},t)|^2\rangle \right)^{\delta} 
= C\langle\|\phi\|^2\rangle^{\delta} \mbox{,}
\end{equation}
where $C$ is a constant and $\delta$ is the value of the slope that we expect to change from the asymptotic $1$ to asymptotic $0.5$ when the transition from weak to strong turbulence takes place. Such a transition in the value of $\delta$, however, is not observed in Fig.~\ref{fig_Q_phi_or}. Instead the data is consistent with a single-valued exponent of around $0.84$ over a wide range of potential amplitudes spanning nearly three orders of magnitude. The closest fit of the relationship given in Eq.~(\ref{eq_Q_form}) to the simulation data is shown in Fig.~\ref{fig_Q_phi_or} by the dashed red line together with the specific numerical values for $C$ and $\delta$.  We note in passing that similar values of $\alpha$ were identified in Ref.~\cite{Hatch2}, wherein turbulent amplitude was mediated by varying background shear flow.  Although this turbulence metric (the scaling of diffusivity with amplitude) has not been widely studied numerically, these results suggest that an intermediate scaling favoring weak turbulence may be characteristic of drift-type plasma turbulence.\\
There is a clear discrepancy between the numerical data and the result expected based on the theory outlined in Ref.~\cite{Zhang1}. If it were not for the deeper understanding (alluded to in the introduction), it would have been a fatal blow to the class of simple analytical theories like Ref.~\cite{Zhang1}.\\
Let us see why in this particular simulation, where the turbulence levels were boosted by increasing $\omega_T$, the system remains stuck to an intermediate value of the exponent $\delta$ without exhibiting any of the asymptotic states indicated by the theory.\\
We were led to anticipate two different asymptotic regimes (WT and ST) by examining the propagator in the expression for $\Pi(\mathbf{k})$ given by  Eq.~(\ref{eq_cap_Pi}). It was contended that the system will approach WT (ST) for $\omega\gg|\mathbf{k}\cdot\mathbf{D}_{jj}\cdot\mathbf{k}|$ ($\omega \ll |\mathbf{k}\cdot\mathbf{D}_{jj}\cdot\mathbf{k}|$), and we tried to engineer the WT-to-ST transition by increasing the turbulent transport via increasing $\omega_T$ and thereby driving the system stronger and stronger (larger growth rates for a wider range of wavenumbers). This expectation depended upon the implicit assumption that, on boosting up the drive, only $|\mathbf{k}\cdot\mathbf{D}_{jj}\cdot\mathbf{k}|$ will increase while $\omega$ remains essentially fixed. {\it This, interestingly, is what did not happen in this simulation of ITG turbulence.} The turbulence levels  do increase with $\omega_T$ (larger effective growth rate $\gamma$), but so does the real frequency $\omega$; the latter is also proportional to $\omega_T$. The net result is that the ratio between the linear and the nonlinear parts of the propagator does not change much and we observe no transition even when fluctuation levels go up by several orders of magnitude. In fact a deeper understanding of the theory would have predicted just the results that the numerical simulation yielded.\\
It is important to emphasize here that, lacking a precise mathematical formulation of the propagator and the diffusion coefficients for the reduced gyrokinetic system given by Eq.~(\ref{eq_dna}), we shall attempt to interpret the transition between the different turbulence regimes in reduced gyrokinetic simulations in terms of the criterion provided in Ref.~\cite{Zhang1}. For example, the real frequency $\omega$ appearing in the linear part of the propagator will be identified with the real frequency corresponding to that linear eigenmode of Eq.~(\ref{eq_dna}) which possesses the largest growth rate. Naturally, the turbulence contribution to the renormalized propagator ($|\mathbf{k}\cdot\mathbf{D}_{jj}\cdot\mathbf{k}|$) will come from the nonlinear evolution of the reduced model, and will be connected to the level of turbulent fluctuations.\\
\begin{figure}[h]
    \centering
    \includegraphics[width=0.8\textwidth]{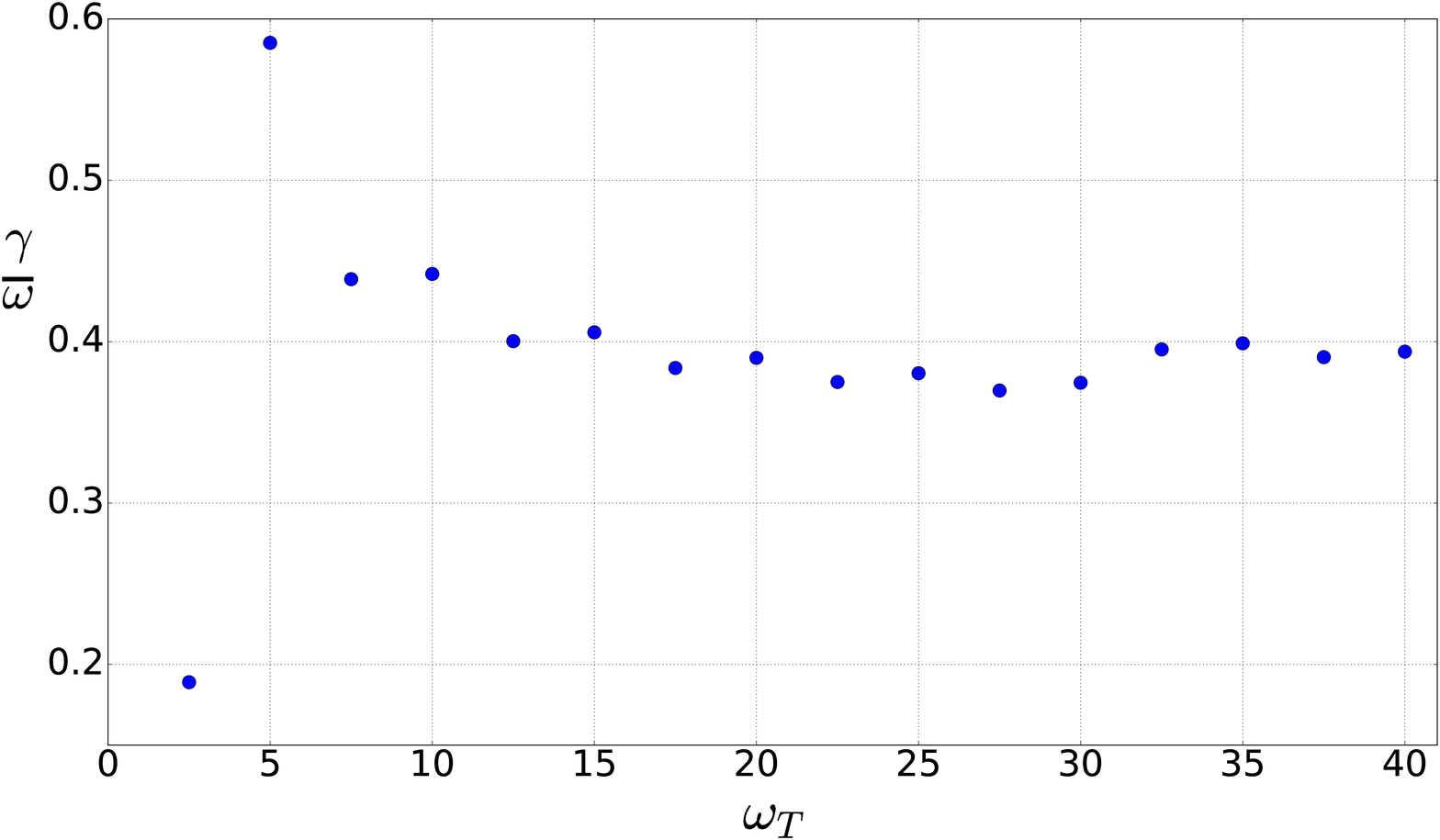}
    \caption{Ratio between growth rate $\gamma$ and real frequency $\omega$ of the most unstable linear mode as a function of the temperature gradient $\omega_T$. For the most part $\gamma/\omega$ is roughly independent of $\omega_T$ meaning that both nominator and denominator scale the same way when the linear drive is increased.}
    \label{fig_g_over_w}
\end{figure}
The kinetic system that we study in this work is driven by internal instabilities and the level of turbulent fluctuations depends on the growth rate of the linearly unstable modes. Hence, consistent with the common mixing-length estimate $D \sim \gamma / k_{\perp}^2$, we expect that the appropriate substitute of $|\mathbf{k}\cdot\mathbf{D}_{jj}\cdot\mathbf{k}|$ in our model will be related to $\gamma$. A subtlety of the kinetic system, however, is that for each $\mathbf{k}$ there are not one but a countable infinity of solutions of the linear part of Eq.~(\ref{eq_dna})~\cite{Hatch_NJP}, i.e., there are infinitely many pairs $(\gamma,\omega)$ to be considered. Nevertheless, for a given wavenumber there is at most one linearly unstable mode that drives the system. As an alternative, then, one could expect $\gamma/\omega$ to serve as a good proxy for $|\mathbf{k}\cdot\mathbf{D}_{jj}\cdot\mathbf{k}|/\omega$. In Fig.~\ref{fig_g_over_w} we have plotted $\gamma/\omega$ as a function of the temperature gradient ($\omega_T$). Since for each $\omega_T$ there are many wavenumbers exhibiting a linearly unstable mode, the $\mathbf{k}$ with the largest growth rate has been selected as representative of the system. It is evident that, for the most part, the ratio $\gamma/\omega$ is roughly independent of $\omega_T$, i.e., both $\omega$ and $\gamma$ scale the same way when the temperature gradient is increased. In addition, a plot of $\gamma$ and $\omega$ alone as functions of $\omega_T$ shows that this scaling is for the most part linear. The first two data points are an understandable exception since in the range of $\omega_T \lesssim 2.2$ there are no linear instabilities, i.e., for $\omega_d = 1$ and $\nu = 0.005$ $\gamma$ changes sign at $\omega_T \approx 2.2$ while $\omega$ does not. Hence, near the onset of the linearly unstable regime it is likely that $\gamma$ and $\omega$ will have a different dependence on the driving parameter $\omega_T$ at least over a very small range of $\omega_T$. The picture seen in Fig.~\ref{fig_g_over_w} is rather robust and does not depend sensitively on which unstable wavenumber we choose as long as its corresponding growth rate is not marginal, but, instead, is comparable to that of the most unstable wavenumber. The plateau in the ratio $\gamma/\omega$ conveys the same information as Fig.~\ref{fig_Q_phi_or} but in a different/simpler way; it could also be seen as a possible explanation why simulations showed no transition in the scaling exponent of $\langle Q\rangle$ versus $\langle\|\phi\|^2\rangle$.
\section{Decoupling of growth rate and real frequency}\label{coll_scan}
\begin{figure}[h]
    \centering
    \includegraphics[width=0.8\textwidth]{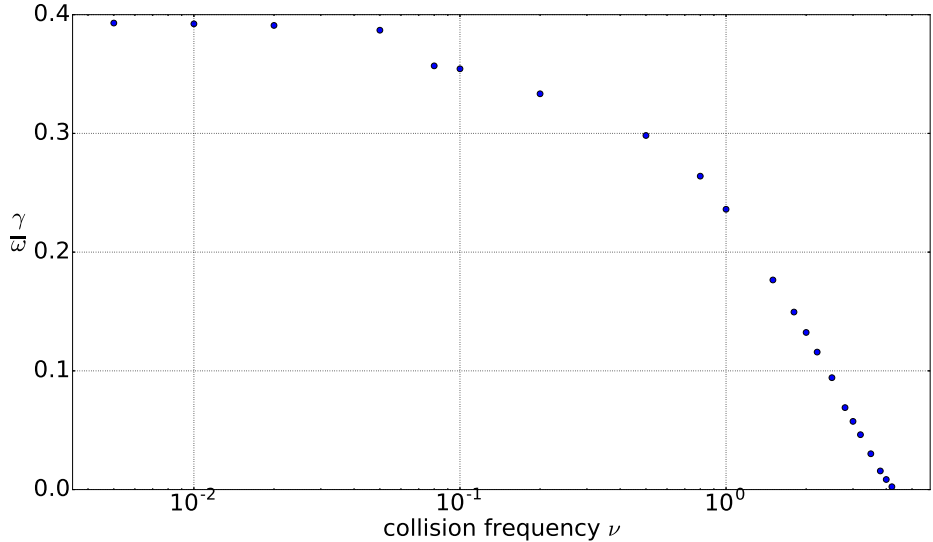}
    \caption{The ratio between growth rate $\gamma$ and real frequency $\omega$ of the most unstable linear mode as a function of collision frequency $\nu$ when $\omega_T = 40$ and $\omega_d = 1$. There is a substantial range, of roughly two orders of magnitude in $\nu$, for which $\gamma/\omega$ shows a clear dependence on collision frequency. (Plot is semi-logarithmic for a better accentuation of the functional variation.)}
    \label{fig_g_over_w_nu}
\end{figure}

\begin{figure}[h]
    \centering
    \includegraphics[width=0.8\textwidth]{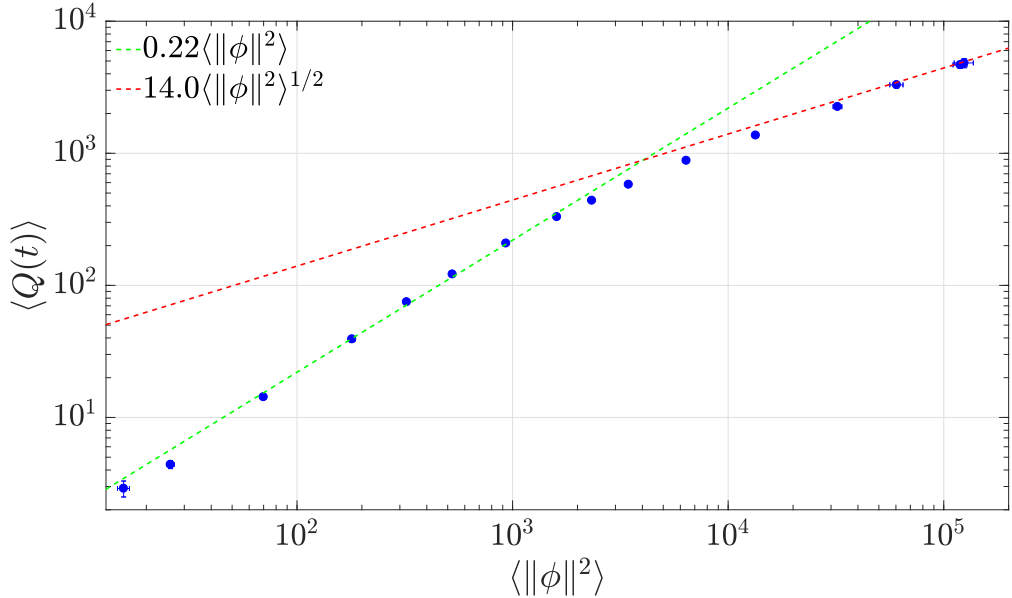}
    \caption{Heat flux as a function of the electrostatic potential (double logarithmic representation) when collision frequency is varied. The two expected asymptotic regimes clearly emerge for both small and large $\|\phi\|$.}
    \label{fig_Q_phi_col}
\end{figure}

\noindent Let us sum up the  results from our first set of simulations. Despite the predictions of the $2$D fluid model in Ref.~\cite{Zhang1} (a renormalized theory of $2$D electrostatic turbulence), numerical simulations failed to show a change in the scaling of heat flux  versus electrostatic potential as its amplitude is increased  with the idea of taking the system from a state of weak (WT) to strong (ST) turbulence. This exponent change in the curve $\langle Q\rangle$ versus $\|\phi\|^2$ was supposed to occur  because at larger levels of turbulence, the second term in the renormalized propagator was expected to become dominant. However, a deeper enquiry into the system provided a rather straightforward explanation for the lack of transition in the kinetic system under study. It did not happen because the $\omega_T$-drive, that we used to push the system harder towards ST, increased also the linear part of the propagator. This crucial feature was strongly emphasized in Fig.~\ref{fig_g_over_w} that shows a plateau in the $\gamma/\omega$ - $\omega_T$ graph.\\
It is rather encouraging to note that, when properly interpreted, the simple analytical model would have predicted exactly the behavior we observed in the given class of kinetic simulations. Now we are ready to proceed with our original quest by, first, posing the question: When and how, if at all, will we observe a change in the exponent $\delta$ (change from weak- to strong-turbulence regime)? Obviously we must choose a setting for the simulations where the growth rate is decoupled from the real frequency, i.e., vary a parameter that increases the linear growth rates (and, therefore, the turbulence levels) but does not influence the real frequency much.\\
The reduced gyrokinetic system (Eq.~(\ref{eq_dna})) is controlled by only three physical parameters (when $\tau$ is fixed): temperature and density gradients ($\omega_T$ and $\omega_d$, respectively), and collision frequency $\nu$. As already observed in Fig.~\ref{fig_g_over_w}, varying $\omega_T$ over a large range yields a plateau for the ratio $\gamma/\omega$. The linear properties of the system dictate similar response to varying $\omega_d$. Therefore, the only natural way of demonstrating a change in the asymptotic slope of the  $\langle Q\rangle$ - $\langle\|\phi\|^2\rangle$ curve is through the variation of the collision frequency. For this ITG system, an increase in collision frequency affects the possible linear instabilities in two ways: 1) the growth rate is decreased, and 2) the instability region in $\mathbf{k}$-space also shrinks. Given any fixed values of $\omega_T$ and $\omega_d$ that allow linear instabilities, there is a threshold in $\nu$ above which all $\mathbf{k}$-modes are stable. For an explicit calculation,  for $\omega_T = 40$ and $\omega_d = 1$ (which sets the stability threshold at $\nu \approx 4.2$), we display in Fig.~\ref{fig_g_over_w_nu}  
a plot of the ratio $\gamma/\omega$ versus $\nu$. 
Notice the palpable contrast with Fig.~\ref{fig_g_over_w}: instead of a plateau, we have a strong variation over much of the range.\\
When collision frequency is small (the left most part of the graph), $\gamma$ and $\omega$ change little because they have well-defined values in the collisionless limit. Although adding a realistic collisional term (like the Lenard-Bernstein collision operator) acts as a singular perturbation to the linear spectrum as a whole\cite{Ng}, the instability itself (if present) changes smoothly with $\nu$ when the latter changes from $0$ to a non-zero value. Further increase in $\nu$ strongly affects the growth rates of all instabilities reducing them to zero at some collisional threshold. However, the real frequency of the unstable mode remains only weakly affected by $\nu$ even at large values of the latter. Hence, the ratio $\gamma/\omega$ smoothly decreases from its collisionless value to zero when collisions become more prominent as evident in Fig.~\ref{fig_g_over_w_nu}. Thus, we seem to have hit an ideal system where we could expect to actually access and distinguish  between the two asymptotic regimes of weak and strong turbulence by varying $\nu$. Nonlinear DNA simulations produce the result shown in Fig.~\ref{fig_Q_phi_col} where each data point is derived from the statistically stationary state of the corresponding simulation.\\
Again, in contrast to Fig.~\ref{fig_Q_phi_or}, we observe in the $\langle Q\rangle$-$\|\phi\|^2$ graph two clear segments with (asymptotically) different slopes. Large collision frequency leads to smaller growth rates of the linear instabilities as well as fewer unstable modes which results in smaller amplitudes for the electrostatic potential and lower levels of heat transport (left part of the figure). In this range $\langle Q\rangle$ scales linearly with $\|\phi\|^2$ consistent with our theoretical expectation for the regime of weak turbulence. The green dashed line seems an excellent fit to the actual points representing results of turbulence simulation for large collision frequency (relatively lower amplitudes). When $\nu$ becomes small enough, and the amplitude of the potential increases beyond some level (for this set of parameters $\sim 2\cdot 10^3$), the  points start to consistently stay below the line, gradually diverging from it. Eventually the $\langle Q\rangle$-$\|\phi\|^2$ curve displays a new regime with a new slope where $Q$ is proportional to $\|\phi\|$ instead of $\|\phi\|^2$.\\
The preceding nonlinear turbulent simulations constitute a strong verification of the fundamental qualitative prediction of a generic renormalized turbulence model (Ref.~\cite{Zhang1}). There are two asymptotic regimes in which a representative turbulent diffusivity scales differently - as $\|\phi\|^2$ in the weakly-turbulent state and as $\|\phi\|$ in the strongly-turbulent limit.  
\section{Direct modification of growth rates}\label{mod_dna}
\begin{figure}[h]
    \centering
    \includegraphics[width=0.8\textwidth]{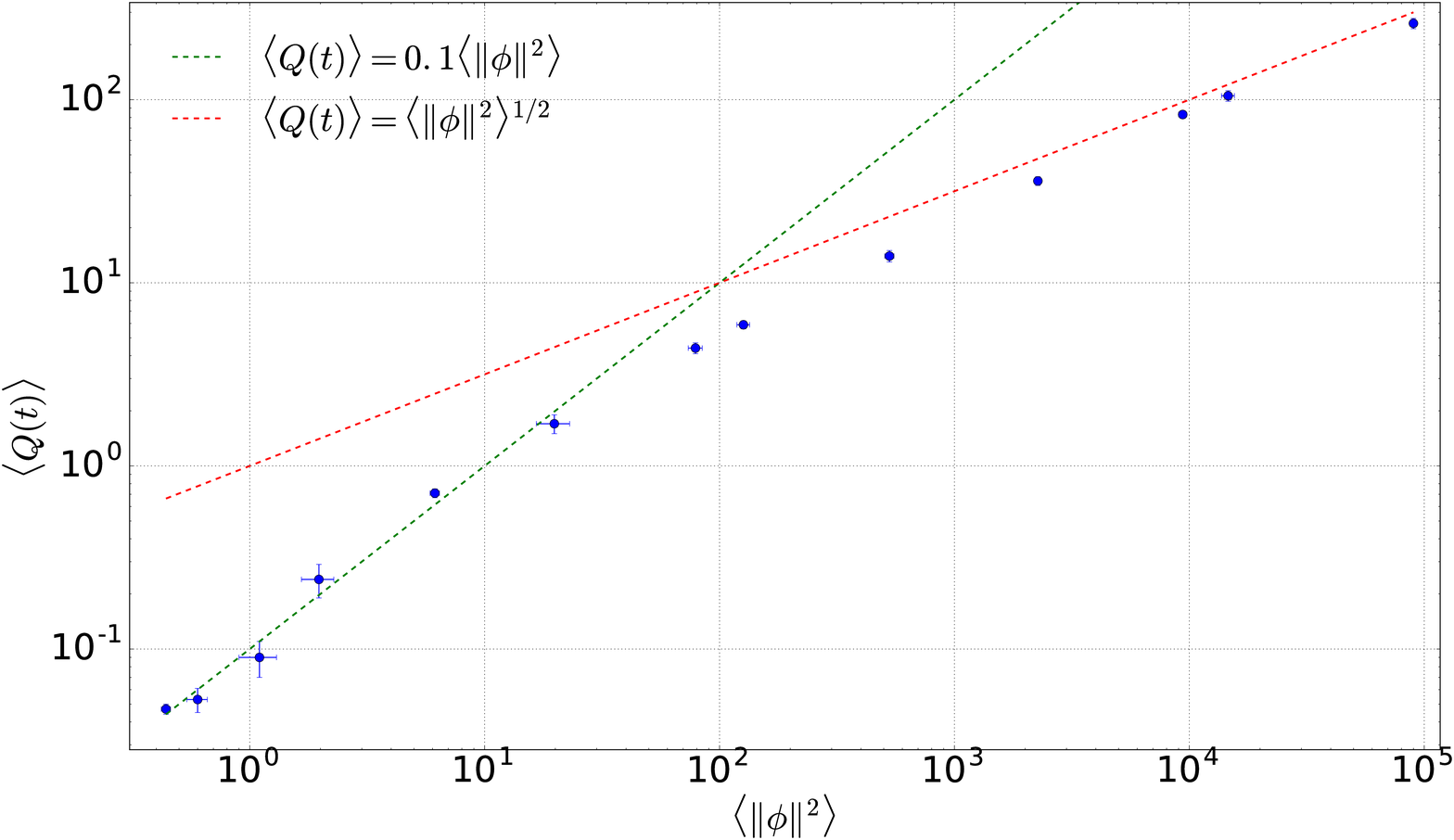}
    \caption{Heat flux as a function of electrostatic potential (double logarithmic representation) when the growth rate of the driving modes is changed directly and independently of their real frequency.}
    \label{fig_Q_phi_mod}
\end{figure}

\noindent In the previous two sections we explored  the possible emergence of two asymptotic regimes in plasma turbulence via nonlinear simulations of a basic system describing a magnetized plasma in slab geometry.  We found that a necessary condition for the system to transition between the two asymptotic regimes is that the ratio $\gamma/\omega$ must also change significantly with the parameter chosen to boost up  the turbulence levels.\\
Strictly speaking, even for a single $\mathbf{k}$, the system described by Eq.~(\ref{eq_dna}) has a countable infinity of linear modes. Hence, infinitely many different ratios of $\gamma/\omega$ can be identified. Then what precisely does $\gamma/\omega$ of Figs.~(\ref{fig_g_over_w}) and (\ref{fig_g_over_w_nu}) mean? We chose $\gamma/\omega$ to correspond  only to the most unstable linear mode of the system since it is the linear instabilities that drive turbulence. However, changing a single parameter, e.g., the collision frequency, simultaneously alters all linear modes. Therefore, since we lack the mathematical form of the renormalized propagator for this kinetic system, it is conceivable that features of other modes, i.e., the drift wave, have also to be taken into account. In this section we shall test for such a possibility by altering directly the linear physics of Eq.~(\ref{eq_dna}) in a way that affects only the linear instability (and more precisely only its growth rate) at a desired $\mathbf{k}$. If we manage to reproduce the same result as in section \ref{coll_scan}, then this will demonstrate that the ratio $\gamma/\omega$ corresponding to the fastest instability of the system is, indeed, the right choice, and its variation with the turbulence level is decisive for transition between turbulence regimes.\\
The details of this somewhat elaborate calculation that is the basis of the simulation results that we present below, are given in Appendix \ref{appen}. Modifying the driving mechanism (so that $\gamma$ is changed while the linear frequency remains unaffected) does, indeed, change the scaling behaviour of the heat flux bringing it in line with the expectations based on Ref.~\cite{Zhang1}. The simulation results displayed in Fig.~\ref{fig_Q_phi_mod}, very similar to those of Fig.~\ref{fig_g_over_w_nu},
clearly show the predicted variation of $\langle Q(t)\rangle$ with $\langle\|\phi\|^2\rangle$. The green asymptote captures the leading behavior of WT with $\langle Q(t)\rangle$ scaling as $\langle\|\phi\|^2\rangle$ while the red asymptote captures the ST regime where $\langle Q(t)\rangle$ scales as $\langle\|\phi\|\rangle$.
\section{Discussion and Conclusions}\label{final}
\noindent This work has probed in depth into potentially universal aspects of turbulence predicted by renormalized turbulence theories by numerically studying electrostatic turbulence in magnetized plasmas.  The most important question that we posed and answered may be succinctly summarized as: How does a typical turbulent effect, like the turbulent diffusivity, change as the turbulence levels are increased?\\
We chose a reduced kinetic model to simulate the turbulent state for a temperature-gradient driven instability, studying the relative scaling of the thermal diffusivity $\langle Q\rangle$ with turbulent fluctuation amplitude $\langle\|\phi\|^2\rangle$.  Interestingly, the simulations maintain a weak turbulence scaling $\langle Q\rangle \sim \langle\|\phi\|^2\rangle$ even at extremely high fluctuation amplitudes.  This scaling is similar to that observed in gyrokinetic simulations mediated by shear flow~\cite{Hatch2} and, we postulate, may be the natural scaling in the common scenario in which mode frequencies and growth rates exhibit similar parameter dependences.  Strikingly, the two asymptotic turbulent states (WT and ST) predicted by a generic turbulence theory can be clearly reproduced by decoupling the linear frequency and growth rate.\\
Although the simulated system is quite different, in detail, from the renormalized turbulence model of Ref.~\cite{Zhang1}, the fact that the simulations capture not only the predicted asymptotic states but also the characteristic exponents in the $\langle Q\rangle$-$\langle\|\phi\|^2\rangle$ relationship, points to the robustness of the turbulence attributes that we have exposed. In fact, one expects that these attributes will define turbulence in even more complicated situations where both the linear as well as the turbulent part of the propagator look quite different.\\
This work identifies yet another manifestation of the persistence of linear physics in plasma turbulence (see, e.g., Ref.~\cite{Hatch_NJP} and references therein).  Moreover, it sheds light on the physics that may underly the effectiveness of quasilinear theory~\cite{Kotsch, Staebler, Bourdelle}, and, importantly, may point to systems in which such theories break down.\\
Finally, it is interesting to note that in the strong turbulence regime the diffusivity has a weaker dependence on the turbulence level. The ST regime is defined when the linear part of the propagator becomes negligible compared to the nonlinear part. Since the linear part (basically the linear frequency) is a measure of the plasma spring constant, i.e, a measure of its ability to resist change (that turbulence tends to induce), it is curious that when the turbulent forces become very strong, the plasma does not quite buckle under this stress (as elastic materials are likely to do) but continues to resist turbulent forces even more strongly than in the state of much lower turbulent stress.
\appendix
\section{Numerical modification of linear growth rates}\label{appen}
\noindent Ignoring the nonlinear terms, one can schematically write Eq.~(\ref{eq_dna}) as 
\begin{equation}\label{eq_lin_L}
\frac{\partial \hat{f}_n}{\partial t} = L[\hat{f}_n] \mbox{,}
\end{equation}
where $L$ denotes the linear operator acting on $\hat{f}_n$. The Hermite coefficients of the distribution function can be combined in a vector of the form $\hat{\mathbf{f}} := (\hat{f}_0, \hat{f}_1,..., \hat{f}_N)$ where $N$ is the last index of the truncation, i.e., $\hat{f}_n \equiv 0$ for $n > N$. There exist more sophisticated truncation schemes\cite{Loureiro} but applying them does not change our results, since we are interested in the frequency and growth rate of the ITG instability. The latter is a fluid mode, i.e., it persists also in a coarse-grained, fluid approximation of this model, and, therefore, its precise properties depend very weakly on the exact truncation at high $n$ as long as it is physically reasonable. One can easily verify that by computing the exact dispersion relation for Eq.~(\ref{eq_lin_L}) and comparing its roots with the matrix eigenvalues to be defined later.\\
With such a vector notation $L$ can be represented as a matrix, say $\mathbf{M}$, acting on $\hat{\mathbf{f}}$ with Eq.~(\ref{eq_lin_L}) becoming
\begin{equation}\label{eq_lin_M}
\frac{\partial \hat{\mathbf{f}}}{\partial t} = \mathbf{M}\cdot\hat{\mathbf{f}} =: \mathbf{q} \mbox{,}
\end{equation}
where $\mathbf{M}$ depends on all the parameters of the system including the wavenumber, i.e., for each $\mathbf{k}$ the matrix has different values. A numerical solution of the linear version of the reduced gyrokinetic system given by Eq.~(\ref{eq_dna}) amounts to computing all eigenvalues and corresponding eigenvectors of $\mathbf{M}$. This is how the data shown in Fig.~\ref{fig_g_over_w} was obtained. For the intended test of our hypothesis we need to modify $\mathbf{M}$ in such a way that only the growth rate of the linear instability at a desired $\mathbf{k}$ will be changed while its real frequency $\omega$ as well as all other eigenvalues (both growth rate and real frequency) remain unaltered. In addition, also the corresponding eigenvectors of all eigenmodes must stay the same, including that of the instability whose growth rate we modify. That way we keep fidelity of the ITG mode altering the physical features of the system as little as possible. Earlier in this work we referred to $\omega_T$ as the `drive' of the system. Mathematically, however, the actual drive are the linear instabilities (driven in turn by $\omega_T$) that enhance the amplitudes of certain modes. Eventually a threshold is reached at which nonlinear terms become important introducing coupling and  energy exchange between different wavenumbers. Hence, $\omega_T$ can be viewed as merely a tool to change the growth rate of those instabilities. The alteration that we shall undertake will change that growth rate directly and in a way that leaves $\omega$ unaffected.\\
Let $\mathbf{v}_{j}$ and $\mathbf{u}_j$ denote, respectively,  the right and left eigenvector of $\mathbf{M}$ ($\mathbf{M}$ is a quadratic 
$(N+1)\times(N+1)$-matrix with unique eigenvalues $\lambda_j$ with $j = 0, 1, ..., N$). We shall construct the matrices $\mathbf{\Lambda}$, $\mathbf{V}$ and $\mathbf{W}$ such that $\Lambda_{ij} = \lambda_j\delta_{ij}$ while $\mathbf{v}_{j}$ constitutes the $(j+1)$th column of $\mathbf{V}$ and $\mathbf{u}_j$ is the $(j+1)$th raw of $\mathbf{W}$. Then the modal decomposition (sometimes also referred to as eigendecomposition) of $\mathbf{M}$ says that
\begin{equation}\label{eq_M_decomp}
\mathbf{M} = \mathbf{V}\cdot\mathbf{\Lambda}\cdot\mathbf{W}^T = \sum_{j=0}^{N}\lambda_j \mathbf{v}_j\otimes\mathbf{u}_j \mbox{.}
\end{equation}
Without loss of generality the eigenvalues can be numbered such that the instability corresponds to $\lambda_0$. In view of the above decomposition the needed change of the linear physics can be accomplished by computing all the eigenvalues of $\mathbf{M}$ with their corresponding left and right eigenvectors, replacing $\lambda_0$ by its new value $\tilde{\lambda}_0 = \omega + \im\tilde{\gamma}$ ($\tilde{\gamma}$ is the desired growth rate), and then assembling $\mathbf{M}$ again in accordance with Eq.~(\ref{eq_M_decomp}). Since the matrix does not depend on time, it is sufficient to do this procedure only once at the beginning of the computation that solves Eq.~(\ref{eq_dna}). However, the DNA code, by which the numerical solution of Eq.~(\ref{eq_dna}) is obtained, does not work with the matrix explicitly but instead constructs the right hand side of the equation directly. Hence, we have to modify $\mathbf{q}$ accordingly. This can be done by using the orthogonality of left and right eigenvectors corresponding to different eigenvalues. After the appropriate normalization one can write that $\mathbf{v}_i\cdot\mathbf{u}_j = \delta_{ij}$. This allows us to filter out from $\mathbf{q}$ the part that corresponds to the instability and modify it. Denoting the modified version by $\tilde{\mathbf{q}}$ we have that
\begin{equation}\label{eq_filter}
\tilde{\mathbf{q}} = \mathbf{q} - \lambda_0 (\hat{\mathbf{f}}\cdot\mathbf{u}_0)\mathbf{v}_0 + 
\tilde{\lambda}_0 (\hat{\mathbf{f}}\cdot\mathbf{u}_0)\mathbf{v}_0 = 
\mathbf{q} + (\tilde{\lambda}_0 - \lambda_0) (\hat{\mathbf{f}}\cdot\mathbf{u}_0)\mathbf{v}_0 \mbox{.}
\end{equation}
The above operation is mathematically equivalent to altering the matrix $\mathbf{M}$ and can be embedded in numerical tools like DNA or GENE but has the disadvantage that it needs to be performed at every time step since $\hat{\mathbf{f}}$ is time dependent. Strictly speaking, it is not necessary that all eigenvalues of $\mathbf{M}$ are different for Eq.~(\ref{eq_filter}) to be applicable. It is simply sufficient if the eigenvalue that we want to modify is nondegenerate, i.e., different from the others. Then its left eigenvector will be orthogonal to the subspace spanned by the right eigenvectors of all the other eigenvalues and the filter technique in Eq.~(\ref{eq_filter}) can be applied.
\begin{acknowledgements}
\noindent The first author of this paper has received financial support for his research by the Deutsche Forschungsgemeinschaft (German Research Foundation).\\
This work was supported by US DOE Contract No. DE-FG02-04ER54742.
\end{acknowledgements}
%

%
\end{document}